\documentclass[journal]{IEEEtran}

\usepackage{comment}
\usepackage{mathtools}
\usepackage{amsmath}
\usepackage{graphicx}
\usepackage{epstopdf}

\usepackage{hyperref} 
\usepackage{cite}
\usepackage{amsmath}
\usepackage{color}

\usepackage{commath}

\begin{document}

\title{Bayesian Integration of Multi-resolutional Grid Codes for Spatial Cognition}

\author{Taiping Zeng, XiaoLi Li, and Bailu Si
\thanks{T. Zeng is with Institute of Science and Technology for Brain-Inspired  Intelligence, Fudan University, Shanghai, China and Key Laboratory of Computational Neuroscience and Brain-Inspired Intelligence (Fudan University), Ministry of Education, China (e-mail:zengtaiping.ac@gmail.com).}
\thanks{X. Li, State Key Laboratory of Cognitive Neuroscience and Learning and IDG/McGovern Institute for Brain Research, Beijing Normal University, Beijing, China (e-mail:xiaoli@bnu.edu.cn).}
\thanks{B. Si is with School of Systems Science, Beijing Normal University, 100875, China (e-mail:bailusi@bnu.edu.cn).}
\thanks{Correspondence should be addressed to Bailu Si (bailusi@bnu.edu.cn).}
}


\maketitle

\begin{abstract}
Fourier-like summation of several grid cell modules with different spatial frequencies in the medial entorhinal cortex (MEC) has long been proposed to form the contours of place firing fields.
Recent experiments largely, but not completely, support this theory. Place fields are obviously expanded by inactivation of dorsal MEC, which fits the hypothesis. However, contrary to the prediction, inactivation of ventral MEC is also weakly broaden the spatial place firing patterns.
In this study, we derive the model from grid spatial frequencies represented by Gaussian profiles to a 1D place field by Bayesian inference, and further provide completely theoretical explanations for expansion of place fields and predictions for alignments of grid components. 
To understand the information transform across between neocortex, entorhinal cortex, and hippocampus, we propose spatial memory indexing theory from hippocampal indexing theory to investigate how neural dynamics work in the entorhinal-hippocampal circuit. 
The inputs of place cells in CA3 are converged from three grid modules with different grid spacings layer II of MEC by Bayesian mechanism.
We resort to the robot system to test Fourier hypothesis and spatial memory indexing theory, and validate our proposed entorhinal-hippocampal model.
And then we demonstrate its cognitive mapping capability on the KITTI odometry benchmark dataset.
Results suggest that our model provides a rational theoretical explanation for the biological experimental results. Results also show that the proposed model is robust for simultaneous localization and mapping (SLAM) in the large-scale environment.
Our proposed model theoretically supports for Fourier hypothesis in a general Bayesian mechanism, which may pertain to other neural systems in addition to spatial cognition.
\end{abstract}


%
\IEEEpeerreviewmaketitle

\section{Introduction}
\label{sec:Intro}


\IEEEPARstart{S}{patial} cognition and navigation are critical for the survival of mammals, which give them impressive capabilities to explore and navigate in an unknown environment. An internal mental map-like representation, called cognitive map~\cite{tolman_cognitive_1948}, has long been hypothesized to guide mammals to learn spatial information and perform space-dependent cognitive tasks, like exploration, map building, localization, path planning, etc. 
The following discovery of place cells supports the idea of cognitive map~\cite{okeefe_hippocampus_1971,okeefe_hippocampal_1978}. When a single neuron of the hippocampus is recorded in the brain of freely moving rats, place cell fires when the rat is within a confined region of the environment. Head direction cells (HD cells), discovered in various brain regions, only fire when the head of rats faces a particular direction, which is thought to be as a compass to provide directional information~\cite{taube_head-direction_1990,taube2007head}. 
In the medial entorhinal cortex (MEC), upstream of the hippocampus, grid cells are characterized by multiple firing fields arranged in a strikingly regular, triangular, grid-like pattern spanning the whole environment explored by mammals~\cite{hafting_microstructure_2005}. The spacing of grid-like pattern increases from dorsal to ventral in the MEC to represent environments with different scales~\cite{hafting_microstructure_2005,sargolini_conjunctive_2006}. Obviously, grid cells and place cells in the entorhinal-hippocampal circuits play an essential role in spatial cognition and navigation.

However, the underlying mechanism that place cells are formed from grid cells is yet to be elucidated. It is obviously possible that grid cells provide spatial inputs to generate place firing fields. In the dentate gyrus, i.e. the input region of the hippocampus, the place responses may be formed during the process of self-organized plasticity from the superficial layers of the entorhinal cortex~\cite{rolls2006entorhinal}. Competitive Hebbian learning rules are applied to generate connections from grid cells in layer II of MEC to granule cells in the dentate gyrus~\cite{si2009role,monaco2011modular}. Meanwhile, grid cells in the layer II/III of MEC also project to the CA3 and CA1 region of hippocampus\cite{deng2010new,kitamura2015entorhinal}. The grid cells in the superficial layers of the entorhinal cortex are the main cortex inputs to the hippocampus, which leads researchers to propose that hippocampal place fields may generated through a Fourier synthesis mechanism~\cite{solstad_grid_2006,mcnaughton_path_2006}. Place fields form by linear summation of grid firing patterns with a common central peak, but different grid spacing and orientation. As the wavelength of grid patterns is different, the grid patterns enhance the central peak and cancel others~\cite{moser_place_2015}.

Recently, neurobiological experiments largely, but not completely, support the Fourier hypothesis~\cite{kubie_spatial_2015,ormond_place_2015}. In this hypothesis, the contours of firing fields are molded by the spatial frequencies from several grid cell modules. The experiment is performed by inactivating grid inputs at multiple sites along the dorsoventral axis of MEC, and at the same time, neurons in areas CA3 and CA1 in the dorsal half of the hippocampus are recored~\cite{ormond_place_2015}. According to the prediction of the Fourier hypothesis, inactivation of dorsal MEC should cause place field expansion, whereas inactivation of ventral MEC should cause place field contraction. Experiments show that inactivation of dorsal MEC indeed expanded the place field; however, contrary to the prediction, inactivation of ventral MEC not only is unable to narrow the place field, but also has a weak tendency to broaden the place filed~\cite{ormond_place_2015}. Ormond and McNaughton suggest that inactivation of part of MEC decreases the self-motion signal to drive all grid cells. Thus, inactivation of part of MEC can alway expand the place firing fields. Kubi and Fox propose an alternative explanation that due to the large size of discrete $\sim$1.7-fold steps~\cite{stensola_entorhinal_2012}, the place field is determined by the highest frequency, whereas the low-frequency determines the location of the place field in the high-frequency cycle, but not the shape of the place field~\cite{kubie_spatial_2015}. 

Still, there is no conclusive explanation about the place field expansion in both the inactivation of the dorsal and ventral part of MEC yet. Also, there lacks theoretical supports for this experimental phenomenon based on the Fourier hypothesis.

In this paper, following anatomical connectivity and cognitive functions, we model the entorhinal-hippocampal circuit to explore spatial navigation and episodic memory during the process of animal navigation. We first derive the model of place cells from different grid spatial frequencies on a general Bayesian mechanism to explain the expansion of place fields and predict alignments of grid components. Then, to account for the whole organization of spatial navigation and episodic memory, spatial memory indexing theory is proposed to describe the information transform across between neocortex, entorhinal cortex, and hippocampus. We organically present that the MEC receives the input of motion and direction, after path integration, and relay to the hippocampus, when the brain experiences a new episode. The hippocampus serves as an index project to the visual cortex via Lateral Entorhinal Cortex (LEC) to activate stored visual episodic memory. When the brain experiences a familiar episode, the stored visual episodic memory leads to find the index in the hippocampus. Further, the hippocampus projects back to the entorhinal cortex to calibrate the presentation of path integration. Moreover, based on our proposed theory, we model the entorhinal-hippocampal circuit. The robot system is utilized to test Fourier hypothesis and spatial memory indexing theory. We validate our entorhinal-hippocampal model on the KITTI odometry benchmark dataset. The results demonstrate the cognitive mapping capability of our system in a large-scale environment. It also further provides beneficial supports for the Fourier hypothesis by Bayesian inference.

The contribution of this paper is three-folded.
First, we strongly support the Fourier hypothesis on a general Bayesian mechanism to explain the expansion of place fields and predict the alignment of grid components. Our derived theoretical model can explain dorsal parts of the MEC causes a broadening of the spatial tuning of place cells; also, inactivating the ventral MEC produces a weak tendency to broaden their spatial firing patterns of place cells~\cite{ormond_place_2015}. We also validate that the high-frequency grid patterns are able to create activity bump on the broad low-frequency bump. The location of the firing field is determined by the low-frequency components, but the shape of the place field is determined by the high-frequency components~\cite{kubie_spatial_2015}. Also, the Fourier-like summation of inputs must align the grid inputs firstly~\cite{mcnaughton_path_2006, solstad_grid_2006}. Our Bayesian inference allows a little shift during alignment of grid inputs. The terrible alignment would degenerate the place field, but our model is tolerant for minor shift and still provides necessary information for spatial cognition.
Second, we propose the spatial memory indexing theory to interpret the intrinsic organization of spatial navigation and episodic memory as a whole, and describe the information transform across between neocortex (visual cortex and vestibular cortex), entorhinal cortex (medial entorhinal cortex and lateral entorhinal cortex), and hippocampus (CA3 and CA1). 
Third, our proposed entorhinal-hippocampal model is successfully implemented and demonstrated for robot navigation in the large-scale environment, based on the benchmark datasets.

\section{Fourier Hypothesis by Bayesian Inference}
\label{sec:fourierBayes}

According to experimental results and further interpretations~\cite{ormond_place_2015,kubie_spatial_2015}, we propose a computational model to parse contradiction about the prediction of hypothesis. We further improve the Fourier hypothesis that place firing fields can be molded from the spatial frequencies of grid cells by Bayesian inference, not by summation~\cite{solstad_grid_2006}.  

\subsection{Derivation}

Grid cells are able to generate remarkable triangular lattice patterns in the explored environment~\cite{hafting_microstructure_2005}. Each grid module can be described by its grid scale, orientation, and phase. The grid scale is thought to be the reciprocal of the grid spatial frequency. The grid scales increase in discrete steps from dorsal to ventral of MEC~\cite{stensola_entorhinal_2012}. 

\begin{figure}[!ht]
\centering
\includegraphics[width=3in]{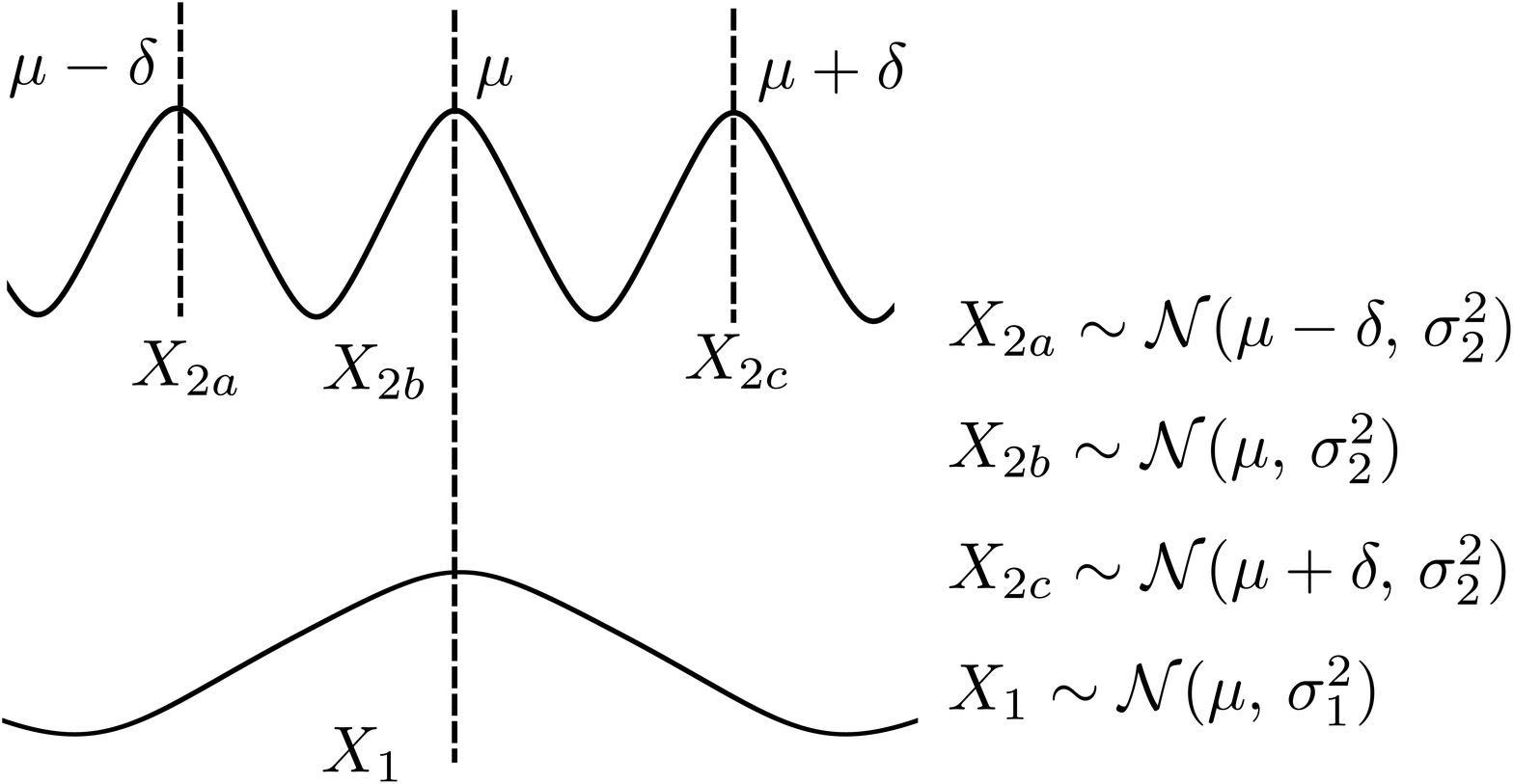}
\caption{Bayesian Inference from spatial grid frequencies with Gaussian profiles to 1D place field.}
\label{fig_bayesFourierDerivation}
\end{figure}

\begin{figure*}[!ht]
\centering
\includegraphics[width=5in]{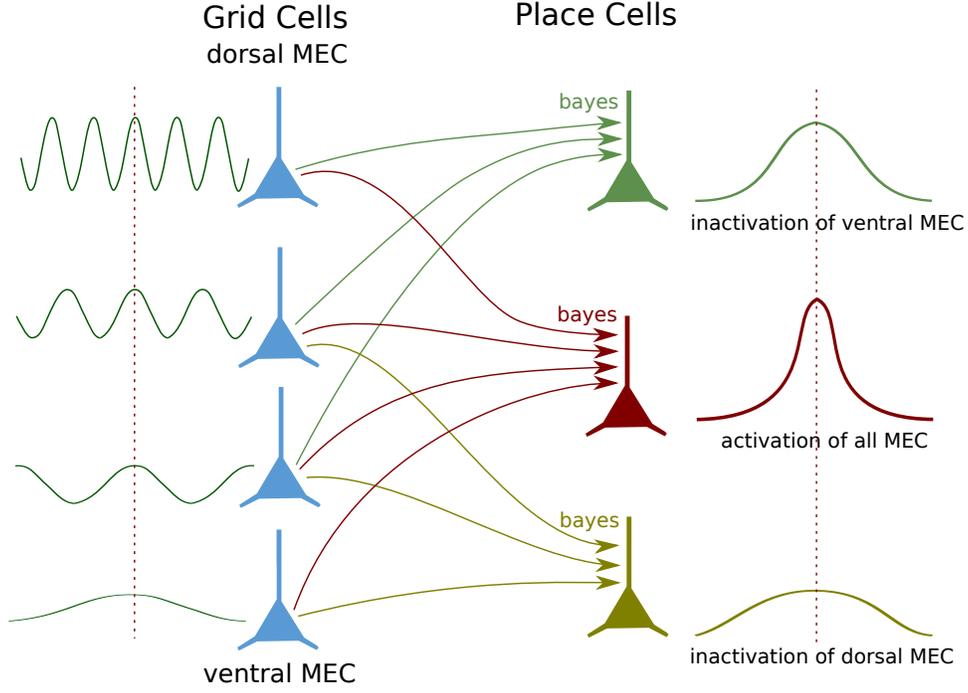}
\caption{Interpretation of place field expansion after inactivation of MEC. Contour of a 1D place field produced by integrating periodic Gaussian profiles.}
\label{fig_bayesFourierExpansion}
\end{figure*}

Here, we simplify 2D grid firing patterns to 1D periodic firing patterns for the explanation of the generation of the place field. Two 1D periodic firing patterns are shown in Fig.~\ref{fig_bayesFourierDerivation}. The bottom panel shows a simple bump with a larger grid scale corresponding to the ventral of MEC, called ventral firing pattern. The ventral firing pattern is supposed to satisfy the Gaussian profile given by
\begin{equation}
\begin{split}
\displaystyle
f(x | \mu, \sigma_1^2) = \frac{1}{\sqrt{2\pi\sigma_1^2}} \exp(-\frac{(x-\mu)^2}{2\sigma_1^2}).
\end{split}
\end{equation}
The ventral firing pattern can be considered to be an encoding of the current locations of the rodent in the neural manifold. The ventral firing pattern describes an estimation of the current location by the activity bump in one grid module, in which $\mu$ is the center of the bump, and $\sigma_1$ describes the broadness of the bump.

The firing pattern in the top panel is with smaller grid scale corresponding to the dorsal of MEC, termed dorsal firing pattern. The dorsal firing pattern is described by the superposition of three Gaussian profiles, which can be defined as
\begin{equation}
\begin{split}
\displaystyle
f(x | \mu - \delta, \sigma_2^2) + f(x | \mu, \sigma_2^2) + f(x | \mu + \delta, \sigma_2^2).
\end{split}
\end{equation}
This dorsal firing pattern can be thought to estimate the current location of the rodent in three potential locations $\mu - \delta$, $\mu$, and $\mu + \delta$ with same probability described by $\sigma_2$.

For further exploring generation of the place field, we propose an interaction between two grid modules similar in Bayesian manner. For the ventral firing pattern, the distribution can be denoted as $p(g_1 | \theta)$, which means the estimation by grid pattern number $1$ given a location $\theta$. For the dorsal firing pattern, three different location distributions can be denoted as $p(g_{2a} | \theta)$, $p(g_{2b} | \theta)$, $p(g_{2c} | \theta)$. For each location estimation, there is a Bayesian interaction with other location estimation in different grid modules. After interaction, in Fig.~\ref{fig_bayesFourierDerivation}, three temporal new Gaussian distributions, the probability density of $p(g_{12a} | \theta) = p(g_{1} | \theta)p(g_{2a} | \theta)$, $p(g_{12b} | \theta) = p(g_{1} | \theta)p(g_{2b} | \theta)$, $p(g_{12c} | \theta)=p(g_{1} | \theta)p(g_{2c} | \theta)$ can be respectively described by
\begin{equation}
\begin{split}
\displaystyle
&f(x | (\frac{\mu}{\sigma_1^2} + \frac{\mu - \delta}{\sigma_2^2})\,\frac{\sigma_{1}^{2}\sigma_{2}^{2}}{\sigma_{1}^{2} + \sigma_{2}^{2}}, \frac{\sigma_{1}^{2}\sigma_{2}^{2}}{\sigma_{1}^{2} + \sigma_{2}^{2}}), \\
&f(x | (\frac{\mu}{\sigma_1^2} + \frac{\mu }{\sigma_2^2})\,\frac{\sigma_{1}^{2}\sigma_{2}^{2}}{\sigma_{1}^{2} + \sigma_{2}^{2}}, \frac{\sigma_{1}^{2}\sigma_{2}^{2}}{\sigma_{1}^{2} + \sigma_{2}^{2}}), \\
&f(x | (\frac{\mu}{\sigma_1^2} + \frac{\mu + \delta}{\sigma_2^2})\,\frac{\sigma_{1}^{2}\sigma_{2}^{2}}{\sigma_{1}^{2} + \sigma_{2}^{2}}, \frac{\sigma_{1}^{2}\sigma_{2}^{2}}{\sigma_{1}^{2} + \sigma_{2}^{2}}).
\end{split}
\end{equation}
Furthermore, the posterior distribution of location phase is given by
\begin{equation}
\begin{split}
\displaystyle
p(\theta | g_{12}) = p(g_{12a} | \theta)\,p(g_{12b} | \theta)\,p(g_{12c} | \theta)\,p(\theta),
\end{split}
\end{equation}
where, $p(\theta | g_{12})$ is probabilistic distribution of location given two grid firing patterns. Considering there is no prior knowledge available, $p(\theta)$ is uniform. The posterior probability can be rewritten as 
\begin{equation}
\begin{split}
\displaystyle
p(\theta | g_{12}) = p(g_{12a} | \theta)\,p(g_{12b} | \theta)\,p(g_{12c} | \theta).
\end{split}
\end{equation}

Considering the number of activity bumps as a scaler factor to sharpness of the place firing field, the probabilistic density of the posterior probability $p(\theta | g_{12})$ can be further described by 
\begin{equation}
\begin{split}
\displaystyle
&f(x | \mu, \frac{\sigma_1^2 \sigma_2^2}{\sigma_1^2 + \sigma_2^2}).
\end{split}
\label{eq:bayesFourierIntegration}
\end{equation}
The final posterior probability $p(\theta | g_{12})$ is thought to describe the place field from the two patterns, namely, the ventral pattern and the dorsal pattern.

Given there exist multiple bumps to represent the current location, a scaler equal to the number of bumps is added to normalize the bump sharpness of place cells. The sharpness of the place field is described by the reciprocal of the scaled variance, i.e., $\displaystyle \frac{1}{\sigma_1^2} +\frac{1}{\sigma_2^2}$. 
Actually, the sharpness of grid patterns in the dorsal pattern is highly larger that the sharpness in the ventral pattern. The sharpness of the place field is largely determined by $\displaystyle \frac{1}{\sigma_2^2}$, and the center of the activity bump is determined by $\mu$.

\subsection{Expansion of Place Fields}

\begin{figure}[!ht]
\centering
\includegraphics[width=2.0in]{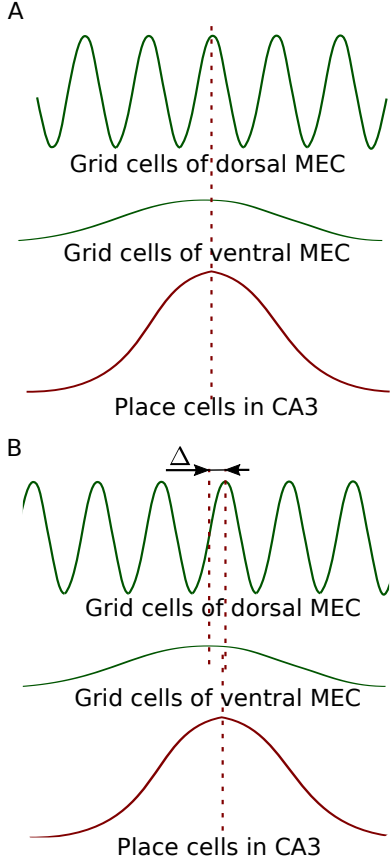}
\caption{Place fields generated by grid alignments through Bayesian inference.}
\label{fig_bayesFourierAlignment}
\end{figure}

Following the explanation of the place firing fields forming from the grid spatial frequencies~\cite{kubie_spatial_2015}, we further explain the results not consistent with the prediction of the hypothesis. As described in the results, inactivating the dorsal part of the MEC strongly, reliably broadens the spatial tuning of place cells; however, inactivating the ventral part weakly causes a broadening of the spatial tuning of place cells. 

We firstly interpret the place field expansion after inactivation of MEC in~\cite{ormond_place_2015}, shown in Fig.~\ref{fig_bayesFourierExpansion}. From ventral to dorsal of MEC, the broadness of grid frequencies can be described by Gaussian standard deviation, i.e. $\sigma_1, \sigma_2,\sigma_3,\sigma_4$, and the broadness of grid spatial frequencies also meets the condition, namely  $\sigma_1 > \sigma_2 > \sigma_3 > \sigma_4$. For the contour of 1D place field with red color, as all parts of MEC are activated, the broadness of the red place field can be described as
\begin{equation}
\begin{split}
\displaystyle
\sigma_{red} = \sqrt{\frac{1}{\frac{1}{\sigma_1^2}+\frac{1}{\sigma_2^2}+\frac{1}{\sigma_3^2}+\frac{1}{\sigma_4^2}}}.
\end{split}
\end{equation}
After ventral MEC inactivation, the ventral spatial frequency of grid cells is missing. Then the broadness of green place field can be described as
\begin{equation}
\begin{split}
\displaystyle
\sigma_{green} = \sqrt{\frac{1}{\frac{1}{\sigma_2^2}+\frac{1}{\sigma_3^2}+\frac{1}{\sigma_4^2}}}.
\end{split}
\end{equation}
The yellow place field loses the dorsal MEC input, given by
\begin{equation}
\begin{split}
\displaystyle
\sigma_{yellow} = \sqrt{\frac{1}{\frac{1}{\sigma_1^2}+\frac{1}{\sigma_2^2}+\frac{1}{\sigma_3^2}}}.
\end{split}
\end{equation}

After dorsal and ventral inactivation of MEC, place spatial firing patterns are broadened. Compared with the activation of all MEC (the red place field), inactivation of the dorsal part of MEC strongly expands the place field firing pattern (the yellow place field), which can be clearly seen from Fig.~\ref{fig_bayesFourierExpansion}. Further, the broadness of the yellow place field ($\sigma_{yellow}$) is obviously larger than the red one ($\sigma_{red}$), since the grid component ($\sigma_1$) with highest spatial frequency is missing. 

For the inactivation of ventral MEC, as the grid component of ventral MEC has the lowest spatial frequency, the green place field is weakly expanded, compared with the activation of all MEC (the red place field). After the ventral grid component ($\sigma_4$) is missing, the green place field ($\sigma_{green}$) is also larger than the red one. However, as the grid frequency of dorsal MEC is larger than the grid frequency of ventral MEC, the green place field ($\sigma_{green}$) with inactivation of ventral MEC is relatively larger than the yellow place field ($\sigma_{yellow}$) with inactivation of dorsal MEC.

To be brief, any parts of MEC are missing, which leads place cells to broaden the place firing field. Since the shape of the place field is largely determined by the high-frequency grid components, inactivation of dorsal MEC (high-frequency grid components) strongly broadens the place firing field, whereas inactivation of ventral MEC (low-frequency grid components) only has a weak tendency to broaden the place field.

\subsection{Alignment of Grid Components}
\begin{figure*}[!ht]
\centering
\includegraphics[width=5in]{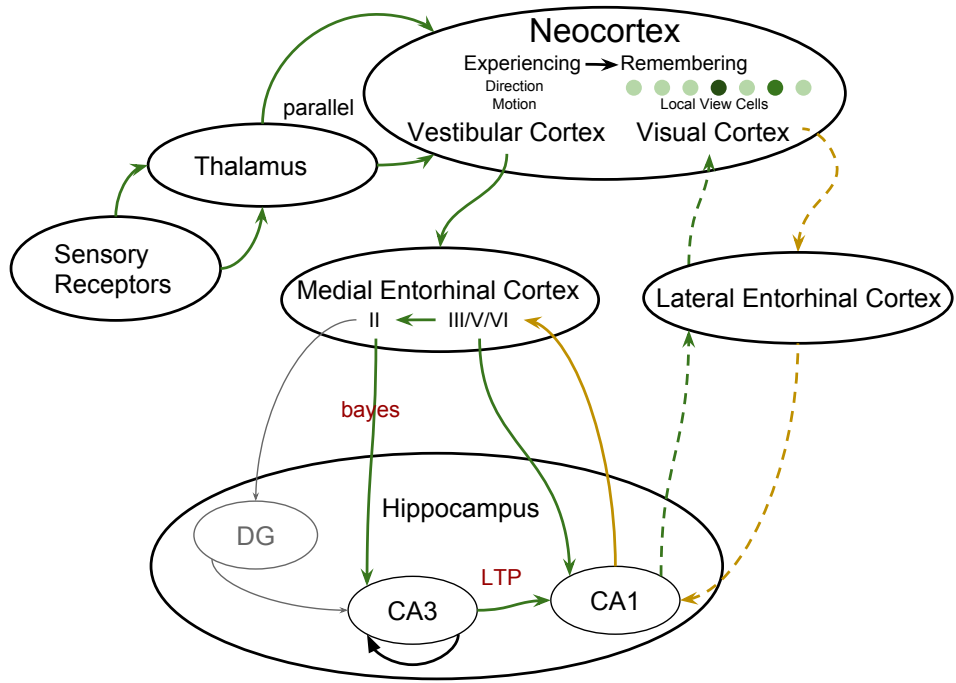}
\caption{Spatial memory indexing theory.}
\label{fig_spatialMemoryIndexingTheory}
\end{figure*}

Since some noises may exist during path integration by grid cells in the MEC, centers of the grid activity bump in all grid modules are very likely to sit in a different position with a relatively small offset. If all grid components are not aligned together, the Fourier-like summation can not completely interpret the formation of the place firing field from grid spatial frequencies. Our computational model is also proposed to predict the alignment of grid frequency components. 

According to our previous model, without offset between two bump centers of grid modules, the place field is generated by Bayesian inference shown in Fig.~\ref{fig_bayesFourierAlignment}A. However, if the bump center of higher grid frequency component is not consistent with the bum center of lower one, the small offset $\Delta$ is existed, shown in Fig.~\ref{fig_bayesFourierAlignment}B. The grid firing pattern of ventral MEC can be described by a Gaussian profile denoted by $f(x | \mu, \sigma_1^2)$. The dorsal firing pattern can be denoted by $f(x | \mu - \Delta, \sigma_2^2)$. Here, the ventral firing pattern is greatly larger than the dorsal one, i.e. $\sigma_1 >> \sigma_2$. According to Bayesian mechanism, the new place firing pattern can be given by 
\begin{equation}
\begin{split}
\displaystyle
f(x | \mu_p, \sigma^2_p),
\end{split}
\end{equation}
where, the broadness of place field is proportional to the Gaussian standard deviation $\displaystyle \sigma_p = \sqrt{\frac{1}{\frac{1}{\sigma_1^2}  + \frac{1}{\sigma_2^2}}}$. The center of place bump activity $\mu_p$ can be given  by
\begin{equation}
\begin{split}
\displaystyle
\mu_p = (\frac{\mu - \Delta}{\sigma_1^2} + \frac{\mu}{\sigma_2^2})\cdot\frac{1}{\frac{1}{\sigma_1^2}  + \frac{1}{\sigma_2^2}}.
\end{split}
\end{equation}
Since $\sigma_1 >> \sigma_2$, $\mu_p$ is determined by $\mu$. Then $\mu_p \approx \mu$.

When all grid frequency components are aligned, shown in Fig.~\ref{fig_bayesFourierAlignment}A, the high-frequency grid component creates a place activity bump on the broad low-frequency peak. In other words, the low-frequency grid component determines which high-frequency cycle creates the place firing field. However, the shape of the place field is determined by the high-frequency grid component~\cite{kubie_spatial_2015}. 
When the bump centers of grid frequency components have a small offset each other, shown in Fig.~\ref{fig_bayesFourierAlignment}B, our proposed model also suggests that the low-frequency grid component determines which high-frequency cycle creates the place firing field and the high-frequency grid component determines the shape of place field. But the precise location of the place field approximates to the peak of the select high-frequency cycle, not the peak of low-frequency grid component.

Simply put, the low-frequency grid component gives the rough location of place field, namely in which high-frequency cycle, whereas the high-frequency grid component determines the precise location (the peak of the selected high-frequency cycle) and the shape of place field.

\section{The Neurobiological Entorhinal-Hippocampal Model}

\subsection{Spatial Memory Indexing Theory}

The hippocampal indexing theory has been proposed to account for episodic memory based on the intrinsic organization, synaptic physiology and cognitive functions of the hippocampus, as well as its anatomical connections to other regions of the brain~\cite{teyler_hippocampal_1986,teyler_hippocampal_2007}. The hippocampus can serve as an index to activate the stored activity patterns in the neocortical regions, and thus retrieve the episodic memory. 

To further illustrate the mechanism of spatial cognition and navigation, we propose the spatial memory indexing theory based on the hippocampal indexing theory, shown in Fig.~\ref{fig_spatialMemoryIndexingTheory}. To simplify this process, we would not further incorporate the pattern separation function of the dentate gyrus (DG) in this research, and focus on the hippocampal pathway (EC-CA3-CA1-EC)~\cite{deng2010new}. 

When the brain experiences an episode, sensory cues are fed forward via thalamus to neocortex in parallel. Different areas of neocortex are activated corresponding to that experience, like the visual cortex for the visual cues in our study. Visual episodic memory is represented by local view cells in the visual cortex. Vestibular cues in the vestibular cortex provide movement information to the medial entorhinal cortex. Path integration is performed by conjunctive grid-by-velocity cells in the layer III, V and VI of MEC. The output layer of MEC, layer II, is projected via the DG onto CA3 in the hippocampus, and layer II of MEC is also directly connected to CA3 region. Then one representation is created as an index in CA3. Neurons in CA3 could perform pattern completion from partial cues through neural attractor dynamics due to the dense reciprocal connections. Layer III of MEC also projects directly to CA1 where a second representation is created~\cite{teyler_hippocampal_2007,deng2010new,kitamura2015entorhinal}. Neurons in CA3 are connected to pyramidal cells in CA1 through schaffer collateral axons. Since neurons in CA3 and CA1 regions are coactivated, their synaptic connections are strengthened through long-term potentiation (LTP). The forward paths are shown as green solid lines in Fig.~\ref{fig_spatialMemoryIndexingTheory}. CA1 also projects back to entorhinal cortex (a yellow solid line in Fig.~\ref{fig_spatialMemoryIndexingTheory}). Simultaneously, many neurons in the primary visual cortex (V1) co-fluctuate with neurons in CA1~\cite{haggerty2015activities}. Synchronized coactivity of CA1 neurons and V1 neurons provides a possibility to selectively strengthen projections from CA1 to visual cortex via lateral entorhinal cortex, shown as a green dashed line in Fig.~\ref{fig_spatialMemoryIndexingTheory}. 

When the brain re-experiences the previous episode, similar areas of neocortex are reactivated. Similar visual cues could reactivate the corresponding local view cells in the visual cortex. Since co-activation between CA1 and visual cortex exists~\cite{haggerty2015activities}, visual cortex may provide inputs to activate the CA1 representation (a yellow dashed line in Fig.~\ref{fig_spatialMemoryIndexingTheory}). When there is a conflict between the activity patterns in CA1 activated from visual cortex and CA3, the consecutive visual episodic memory may gradually influence path integration in the deep layer of MEC through CA1 projections backs to layer III, V, and VI of entorhinal cortex. Thus, the layer II of MEC would project to CA3. And then neurons in CA3 would reactivate neurons in CA1 through connections produced by the LTP process, until the desired activity patterns in CA1 are activated from the visual cortex consistent with from the CA3 region.

\subsection{Neural Network Architecture}

\begin{figure*}[!ht]
\centering
\includegraphics[width=7in]{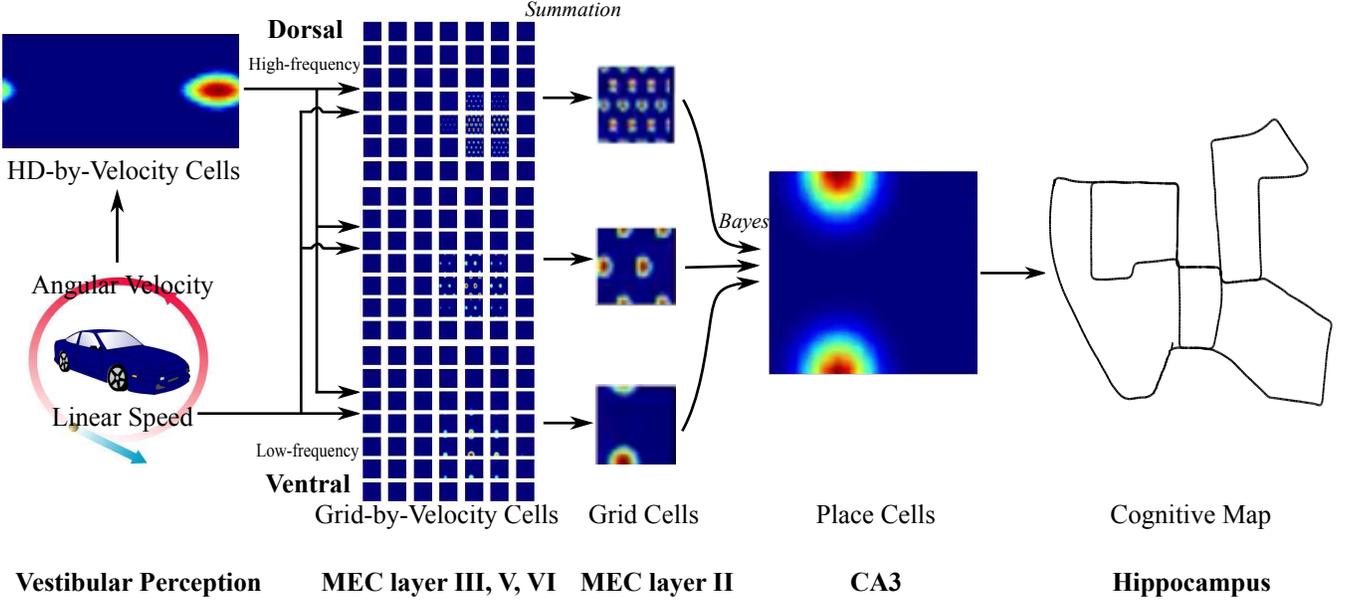}
\caption{The neural network architecture of the model and the diagram of information flow. The HD-by-velocity cells, updated by angular velocity, represent head directions. The same translational velocity inputs of three grid-by-velocity cell network are converted from linear speed and the head direction representations. Three different spatial grid modules from high-frequency to low-frequency correspond to MEC from dorsal to ventral. The activity of grid cells of MEC layer II is formed by the summation of corresponding grid-by-velocity along velocity dimensions. The place cell network in CA3, which is recurrently connected, is driven by the Bayesian integration of three spatial grid frequencies. The positional representation is employed to build a cognitive map.}
\label{fig_bayesFourierSysArch}
\end{figure*}

The neural network architecture of the proposed model is shown in Fig.~\ref{fig_bayesFourierSysArch}. Angular velocity and linear speed are estimated by a proved direct sparse visual odometry (DSO) from a moving stereo camera~\cite{wang_stereo_2017}. 

The head direction of the agent is represented by an activity bump in the ring attractor network of HD cells through population coding. The input of angular velocity could activate a subset of HD cells. The neural attractor dynamics would drive the activity bump to rotate with the same angular velocity as the real input. Due to the periodic boundary conditions, a torus attractor is generated in the conjunctive grid cell network. The translational velocity is converted from linear speed and the head direction representations, which is fed into the grid-by-velocity network to activate a subset of conjunctive grid cells. The grid activity bump travels with a velocity proportional to the movement of the agent in the physical environment through intrinsic attractor dynamics.

Here, three grid modules with different spatial frequencies from high-frequency to low-frequency generate three different grid patterns, corresponding to the MEC from dorsal to ventral~\cite{solstad_grid_2006}. Pure grid cells in layer II of MEC are shaped by removing velocity dimensions of conjunctive grid cells. The inputs of place cell network is provided by Bayesian integration of three grid cells with different spatial frequencies through methods in Section \ref{sec:fourierBayes}. After attractor dynamics, the single activity bump would be formed in the CA3 region of hippocampus to encode the positions of the agent in the physical environment. Thus, the positional representation is utilized to build a cognitive map.

\subsection{Model of HD-by-Velocity Cells}

Angular velocity is integrated by the HD-by-velocity cell network representing a one-dimensional head direction. Head direction and rotation are conjunctively encoded in the HD-by-velocity cell network.

\subsubsection{Neural Representation of Head Direction and Angular Velocity}

Each unit in the network is labeled by its coordinate ($\theta , \nu$) on a two dimensional neural manifold. $\theta \in [0, 2\pi)$ is the internal representation of head directions with periodic boundary condition in the environment. Angular velocity is encoded by $\nu \in [-L_r, L_r]$. Both $\nu$ and $\theta$ are dimensionless quantities in the neural space, which are related to the real angular velocity and the head direction of the robot. The connection weights are designed between units to generate a single stable bump of activity both in the direction dimension $\theta$ and in the rotation dimension $\nu$.
The strength of the connection between a presynaptic unit $(\theta',\nu')$ and a postsynaptic unit $(\theta,\nu)$ can be defined as
\begin{equation}
J(\theta,\nu|\theta',\nu')=J_0+J_1\cos\left(\theta - \theta' - \nu'\right)\cos\left(\lambda (\nu - \nu')\right).
\end{equation}
where $J_0<0$ is a uniform inhibition, $J_1>0$ describes the interaction strength, and $\lambda$ defines the spread of velocity tuning. 
As the center of the postsynaptic unit is not at $\theta'$ but at $\theta'+\nu'$, the connection weight from the unit at $\theta'$ to the unit at $\theta$ in the direction dimension is asymmetric. The asymmetric weights cause the bump to move along the direction dimension with a velocity determined by $\nu'$.

\subsubsection{Network Dynamics}
The bump activity of the network is driven by velocity and sensory inputs. The firing rate $m(\theta,\nu)$ of the unit at coordinate ($\theta , \nu$) can be defined as 
\begin{equation}
\begin{split}
\displaystyle
\tau \dot{m}(\theta,\nu) &= -m(\theta,\nu) \\
&+ f\left(\iint D\theta D\nu J(\theta,\nu|\theta',\nu')m(\theta',\nu') + I_{\nu} + I_{view}\right)\label{eq:HD_dynamics},
\end{split}
\end{equation}
where $I_{\nu}$ and $I_{view}$ are the velocity tuned input and the calibration current injected by local view cells respectively, which we will explain in detail below. $\tau$ is the time constant set to 10 ms. $f(x)$ is a threshold-linear function: $f(x) \equiv [x]_+ = x$ when $x>0$ and 0 otherwise. The shorthand notations are used $\displaystyle \int D\theta = \frac{1}{2\pi}\int_{0}^{2\pi} d\theta$, and $\displaystyle \int D\nu = \frac{1}{2L_r}\int_{-L_r}^{L_r} d\nu$.

\subsubsection{Angular Velocity Inputs}\label{sec:v-input-hd}
To integrate the real angular velocity of the robot, it must be mapped onto the neural manifold of the HD network at an appropriate position. 
Given an external angular velocity $V$, the desired bump location in the $\nu$ dimension can be written as \cite{si_continuous_2014}
\begin{equation}
u(V)=\arctan\left(\tau V \right). \label{eq:angular-velocity-map}
\end{equation}
Here $\tau$ is the time constant defined in Equation~\ref{eq:HD_dynamics}. 
The velocity input to the HD-by-velocity units is simply modeled by a tuning function of Gaussian shape
\begin{equation}
I_{\nu}(\nu|V)=I_r\left[1-\epsilon_r + \epsilon_r,\exp\left(-\frac{(\nu-u(V))^2}{2\sigma_r^2}\right)\right]. \label{eq:angular-v-input}
\end{equation}
Here $I_r$ is the amplitude of the rotational velocity input, $\epsilon_r$ defines the strength of velocity tuning, and $\sigma_r$ is the sharpness of the velocity tuning.

\subsubsection{Estimation of Head Direction}
The head direction and angular velocity of the robot in the physical environment are encoded by the activity bump on the $\theta$ axis and $\nu$ axis. Fourier transformations are utilized to recover the head direction and angular velocity of the robot from the neural activity of the network
\begin{equation}
\psi = \angle(\displaystyle \iint m(\theta, \nu) \exp(i \theta) D\theta D\nu),\label{eq:hd-estimation}
\end{equation}
\begin{equation}
\phi = \frac{\displaystyle \angle(\iint m(\theta, \nu) \exp(i \lambda \nu) D\theta D\nu)}{\lambda}.\label{eq:HD_speed_est}
\end{equation}
Here $i$ is the imaginary unit, and function $\angle (Z)$ takes the angle of a complex number $Z$. $\psi \in [0, 2\pi)$ is the estimated phase of the bump in the direction axis of the neural space, corresponding to the head direction of the robot in the physical environment. $\phi \in (-L_r,L_r)$ is the estimated phase of the bump in the velocity axis of the neural space, and can be recovered to the angular velocity of the robot in the physical space by inverting Equation~\ref{eq:angular-velocity-map} 
\begin{equation}
V = \frac{\tan(\phi)}{\tau}.
\end{equation}
Note that $L_r$ should be selected large enough, so that the recovered velocity $V$ is capable of representing all possible angular velocities of the robot.

\subsection{Model of Grid-by-Velocity Cells}
Then, our HD-by-velocity cell model is expanded to do path integration in the two-dimensional environment. Two-dimensional spatial locations and two-dimensional velocity are represented in the grid-by-velocity cell network. The units in the network are wired with appropriate connection profiles, so that the hexagonal grid firing pattern is created and translated in the spatial dimension of the neural manifold.

\subsubsection{Neural Representation of Position and Velocity}
Units in the grid-by-velocity network is labeled by coordinates $(\vec{\theta},\vec{\nu})$ in a four dimensional neural space.  $\vec{\theta} = (\theta_x,\theta_y)$ represents two dimensional positions with periodic boundary conditions in the environment, i.e. $\theta_x,\theta_y \in [0,2\pi)$. $\nu_x$ and $\nu_y$ are chosen in $[-L_t,L_t]$. $\vec{\nu} = (\nu_x,\nu_y)$ encodes the velocity components in the environment. The connection weights from unit $(\vec{\theta'},\vec{\nu'})$ to $(\vec{\theta},\vec{\nu})$ is described as 
\begin{equation}
\begin{split}
\displaystyle
J(\vec{\theta},\vec{\nu}|\vec{\theta'},\vec{\nu'}) &=J_0 + J_k\cos\left(k\sqrt{\sum \limits_{j\in\{x,y\}}||\theta_j -\theta'_j - \nu'_j ||^2 } \right) \\
&\cos\left(\lambda\sqrt{\sum \limits_{j\in\{x,y\}} (\nu_j - \nu'_j)^2}\right)\label{eq:grid_weight},
\end{split}
\end{equation}
where integer $k = 2$, is chosen so that the network accommodates two bumps both in $\theta_x$ axis and in $\theta_y$ axis. There is only one bump in each of the velocity dimensions however. $||d||$ is the distance on a circle: $||d|| = \textrm{mod}(d + \pi, 2\pi) - \pi$, and $\textrm{mod}(x,y) \in [0,y)$ gives x modulo y. 

\subsubsection{Network Dynamics}
Although the manifold structure of the grid-by-velocity cell network are different with the HD-by-velocity cell network, they share the same intrinsic dynamics as in Equation~\ref{eq:HD_dynamics}
\begin{equation}
\begin{split}
\displaystyle
\tau \dot{m}(\vec{\theta},\vec{\nu}) &= -m(\vec{\theta},\vec{\nu}) \\
&+ f\left(\iint D\vec{\theta} D\vec{\nu}J(\vec{\theta},\vec{\nu}|\vec{\theta'},\vec{\nu'})m(\vec{\theta'},\vec{\nu'}) + I_{\nu} + I_{view}\right).
\end{split}
\end{equation}
Note that $\displaystyle \int D\vec{\theta} = \frac{1}{4\pi^2}\int_{0}^{2\pi}\int_{0}^{2\pi} d\theta_x d\theta_y$, and \newline
 $\displaystyle \int D\vec{\nu} = \frac{1}{4L_t^2}\int_{-L_t}^{L_t}\int_{-L_t}^{L_t} d\nu_x d\nu_y$.

\subsubsection{Translational Velocity Inputs}\label{sec:v-input-grid}
For performing accurate path integration, the input velocity of the robot in the physical environment should be proportional to the velocity of the moving bumps in the neural manifold. The activity bumps are pinned on appropriate positions on the velocity axes according to the tuned velocity inputs, so that the bumps move with the desired velocity.

The translational velocity $\vec{V} = (V_x,V_y)$ of the robot is calculated from the head direction estimated from HD-by-velocity units (Equation~\ref{eq:hd-estimation}) and linear speed by projecting to the axes of the reference frame.
The running speed is encoded by the speed cells in the MEC of the rodent brain. Given the translational velocity of the robot, the desired positions on the velocity axes in the neural space are given by $\vec{u}(\vec{V})$\cite{si_continuous_2014}
\begin{equation}
\vec{u}(\vec{V})=\frac{1}{k}\arctan\left(\frac{2\pi \tau \vec{V}}{S}\right), \label{eq:grid_velocity_input}
\end{equation}
where the function $\arctan$ operates on each dimension of $\vec{V}$. S is a scaling factor between the external velocity of the robot in the physical environment and the velocity of the bumps in the neural space. S determines the spacing between the fields of grid firing pattern in the environment.

The velocity-tuned inputs to the grid-by-velocity units are tuned by a Gaussian form for simplicity
\begin{equation}
I_\nu(\vec{\nu}|\vec{V})=I_t\left[1-\epsilon + \epsilon \,\exp\left(-\frac{|\vec{\nu}-\vec{u}(\vec{V})|^2}{2\sigma_t^2}\right)\right],\label{eq:grid_velocity_tunning}
\end{equation}
where $|\cdot|$ is the Euclidean norm of a vector. $I_t$ is the amplitude of the translational velocity input.    

More detail information about models of HD-by-velocity cells and grid-by-velocity cells can be available in our previous work~\cite{si_continuous_2014, zeng2017cognitive}.

\subsection{Model of Place Cells}

Neurons in the CA3 of the hippocampus are recurrently connected to perform the function of pattern completion. The CA3 region of the hippocampus has been thought to be an attractor network, which attracts firing activities of the network to a stored pattern, even when the external input is incomplete~\cite{renno2014signature}.
We build an attractor network model of place cells in CA3 of the hippocampus. The input of the place cell network is provided by Bayesian integration of three different spatial frequencies of grid cells. 

\subsubsection{Neural Representation of Position}

Each unit in the place cell network is labeled by its coordinate $\vec{\theta}$ in the two dimensional neural manifold. $\vec{\theta} = (\theta_x,\theta_y)$ represents two dimensional positions in the environment, namely $\theta_x,\theta_y \in [0,2\pi)$. The strength of the connection from a presynaptic unit $\vec{\theta}'$ to postsynaptic unit $\vec{\theta}$ can be given by 
\begin{equation}
J(\vec{\theta} |\vec{\theta'} )=J_0+J_1\cos\left(\sqrt{\sum \limits_{j\in\{x,y\}}||\theta_j -\theta'_j||^2 } \right)\label{eq:place_weight},
\end{equation}
where, $||d||$ is the distance on a circle: $||d|| = \textrm{mod}(d + \pi, 2\pi) - \pi$, and $\textrm{mod}(x,y) \in [0,y)$ gives x modulo y.

\subsubsection{Network Dynamics}

The bump activity of the place cell network is only driven by inputs of multiple grid frequencies modules. The place cell network shares the similar intrinsic dynamics with the HD-by-velocity cell network and the grid-by-velocity cell network. The firing rate $m(\vec{\theta})$ of the unit at coordinate $\vec{\theta}$ can be described as
\begin{equation}
\displaystyle
\tau \dot{m}(\vec{\theta}) = -m(\vec{\theta}) + f\left(\int D\vec{\theta} J(\vec{\theta}|\vec{\theta'})m(\vec{\theta'}) + I_{grid}\right),
\end{equation}
where $I_{grid}$ is the input from the Bayesian integration of different grid spatial frequencies. 

\subsubsection{Inputs From Grid Layers}
\label{subsec:inputs_from_grids}
To estimate the center of grid firing activity bump, units belonging to one of activity bumps are clustered.
For each cluster, Fourier transformations are utilized to estimate centers for all activity bumps in the grid cell neural network
\begin{equation}
\begin{split}
\displaystyle
\mu_{\vec{\theta}} = \angle(\displaystyle \int m(\vec{\theta}) \exp(i \vec{\theta}) D\vec{\theta}).\label{eq:bump_center_estimation_mean}
\end{split}
\end{equation}
We also calculate the variance for each activity bump
\begin{equation}
\begin{split}
\displaystyle
\sigma^2_{\vec{\theta}} = \frac{1}{n}\,\displaystyle \int m(\vec{\theta}) (||\vec{\theta} - \mu_{\vec{\theta}}||)^2 D\vec{\theta}, \label{eq:bump_center_estimation_variance}
\end{split}
\end{equation}
where $\norm{d}$ is also the distance on a circle.

In the entorhinal-hippocampal model, we model three grid layers with wave number of activity bumps equaling to 1,2,4 along an axis, respectively. For two-dimensional grid neural manifold with periodic boundary condition, there exist 1,4,16 activity bumps.
After estimation with statistics, we present grid activity patterns as gaussian  profiles. According to our Bayesian inference for spatial frequencies of grid cells molding the firing fields of place cells, the gaussian profile of final input $f(\theta |\mu_{f}, \sigma^2_{f})$ to place cell network in CA3 can be described as
\begin{equation}
\begin{split}
\displaystyle
\frac{1}{\sigma^2_{\vec{\theta}_{f}}} &= \frac{1}{\sigma^2_{\vec{\theta}_{1}}} + \frac{1}{\sigma^2_{\vec{\theta}_{2}}} + \frac{1}{\sigma^2_{\vec{\theta}_{3}}},\\
\frac{\mu_{\vec{\theta}_{f}}}{\sigma^2_{\vec{\theta}_{f}}} &= \norm{\left(\norm{\frac{\mu_{\vec{\theta}_{1}}}{\sigma^2_{\vec{\theta}_{1}}} + \frac{\mu_{\vec{\theta}_{2}}}{\sigma^2_{\vec{\theta}_{2}}}}\right) + \frac{\mu_{\vec{\theta}_{3}}}{\sigma^2_{\vec{\theta}_{3}}}}. \label{eq:place_gaussian_profile}
\end{split}
\end{equation}

Influenced by the intrinsic dynamics of grid attractor network and the inputs from movement and visual sensory, the profile of grid activity bumps are not always circular for $\sigma^2_{\vec{\theta}_{f}} = (\sigma_{\theta_{fx}},\sigma_{\theta_{fy}})$, i.e. $\sigma_{\theta_{fx}}$ not always equals to $\sigma_{\theta_{fy}}$. We separately consider the variance in x and y axes.
Since we set the same number of place units as grid units in each layer, $I_{grid}$ can be described by a Gaussian profile with periodic boundary conditions by sampling 
\begin{equation}
\begin{split}
\displaystyle
I_{grid} = &\frac{1}{\sqrt{2\pi \sigma_{\theta_{fx}} \sigma_{\theta_fy} }} \\
&\exp \left( - \left( \frac{( \norm{\theta_x - \mu_{\theta_fx} } )^2}{{2\sigma_{\theta_fx} ^2 }} + \frac{ ( \norm{\theta_y - \mu_{\theta_fy} } )^2}{{2\sigma_{\theta_fy} ^2 }}\right)\right).\label{eq:input_grid}
\end{split}
\end{equation}


\subsection{Calibration from Visual Cortex}

As the error accumulates during path integration, calibrations from vision are indeed necessary for the positions and head directions of the agent. Local view templates are extracted from the camera images to encode the current scene~\cite{milford_mapping_2008}. If the current local view is novel, a new local view cell in the visual cortex is created. At the same time, the new local view cell is connected to neurons in the CA1 region of the hippocampus through hebbian learning, whose activity patterns are the same as the activity patterns in the CA3 region of the hippocampus in our model. The activity patterns in CA1 also co-activated with grid cells in the deep layer of MEC. Thus, as a new local view cell is created, a local view template, a place activity pattern, the HD activities, and three grid activities are added into the system. If the agent visits a familiar location, the corresponding local view cell is activated. It would inject energy into the HD-by-velocity cell network and the three grid-by-velocity cell networks via place cells in CA1. The HD activities, and the three grid activities stored with the local view cell are scaled and expanded into velocity dimensions with the same values, and then, injected into the corresponding networks. As the continuous familiar local views are fed into the system, the HD and grid patterns would be consistent with the patterns stored with local view cells. Then, the desired activity patterns in CA1 are activated from the visual cortex consistent with the CA3 region.

\section{Implementation of Robot Systems}

\begin{figure}[!ht]
\begin{center}
\includegraphics[width=8cm]{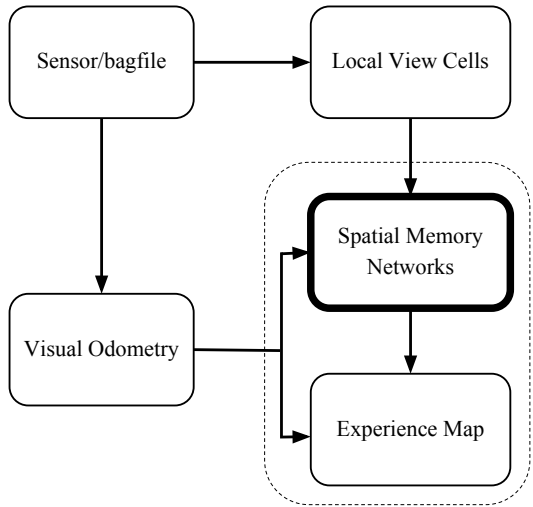} 
\end{center}
\caption{The software architecture of the cognitive mapping system based on the entorhinal-hippocampal circuit. The stereo images are provided by the sensor/bagfile node. Velocity is estimated by visual odometry node. The local view cell node determines whether the current view is novel or not. The spatial memory network node including the entorhinal-hippocampal circuit performs path integration and decision making to create links and vertices. The topological map is built by the experience map node.}\label{fig:node_structure}
\end{figure}

The cognitive mapping system based on our proposed entorhinal-hippocampal model is implemented in Robot Operating System (ROS) Indigo on Ubuntu 14.04 LTS (Trusty) with C++ language. The software architecture of our system is organized into five nodes shown in Fig.~\ref{fig:node_structure}.

The visual odometry node real-time estimates the angular velocity and translational speed based on direction sparse method from a moving stereo camera. It receives images from ROS either from a camera or from stored data in a bagfile. 

The local view cell node determines whether the current image view is a novel or not. It provides calibration current to the networks in the spatial memory node. 

The spatial memory network node organically integrates the entorhinal-hippocampal model, including HD cells, conjunction grid cells in the MEC layer III, V, and VI, grid cells in the MEC layer II, and place cells in CA3 and CA1 of the hippocampus together. This node receives two types of ROS messages as inputs: odometry and view templates. As shown in section~\ref{sec:v-input-hd} and~\ref{sec:v-input-grid}, the HD-by-Velocity cell network and grid-by-Velocity cell network integrate velocity information and visual information to form neural codes. The recurrently connected place cell network takes the Bayesian integration of three grid modules with different spatial frequencies as inputs. The single activity bump would emerge in the place cell network through attractor dynamics to represent the location of the robot in the environment. The spatial memory node also makes decisions about the creation of vertices and links in the experience map, and sends ROS messages of graph operations to the experience map node.

The experience map node builds a coherent cognitive map from the neural codes of the place units. The key locations of the environment are represented as the vertices in a topological graph. A vertex stores the position estimated from the spatial memory network. A link maintains odometric transition information between vertices. On loop closure, a topological map relaxation method is used to find the minimum disagreement by optimizing the positions of vertices~\cite{duckett2002fast}. When the current position in the spatial memory network is far enough from the position of the previous vertex, a new vertex is created and a new edge is connected to the previous vertex.

\subsection{Neural Representation}
\begin{figure*}[!ht]
\begin{center}
\includegraphics[width=13cm]{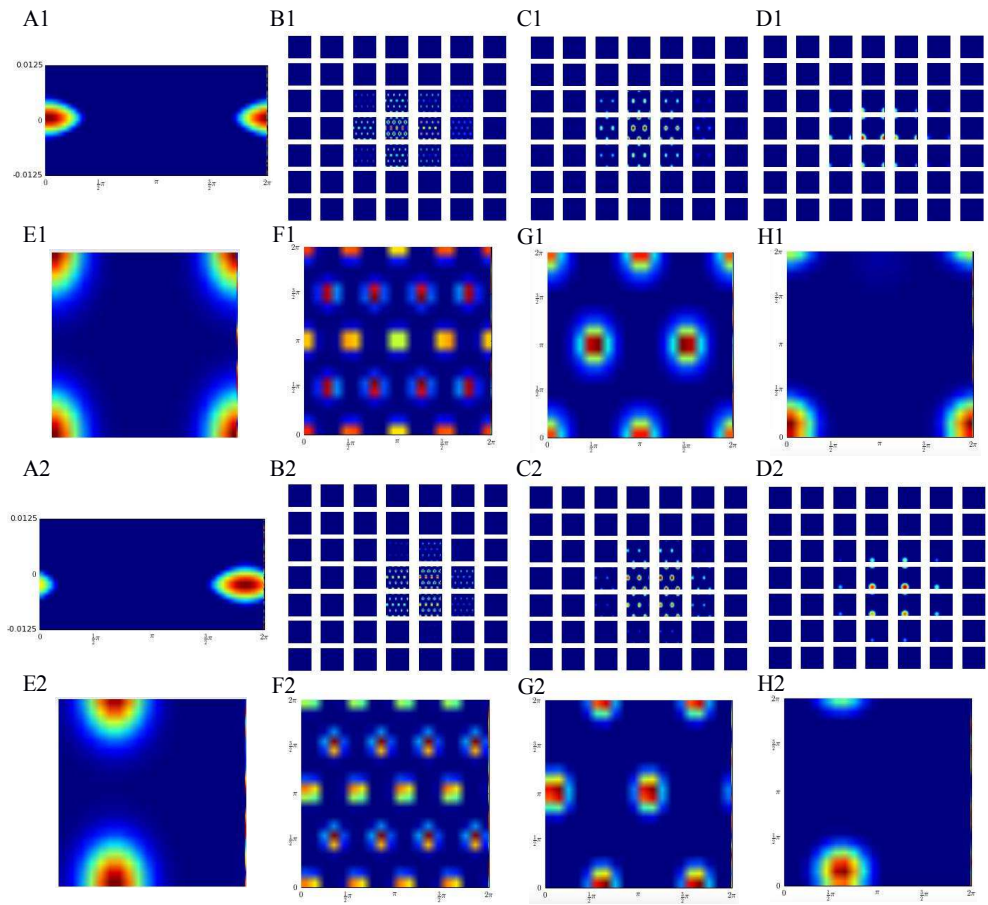} 
\end{center}
\caption{Neural representations in the cognitive mapping model. The activity of each unit from zero to maximal is color-coded from blue to red. Two groups of figures show activities of the units in the beginning of the experiments when the robot is static, and at a randomly selected time stamp when the robot is moving. (A) The population activity of the HD-by-velocity units. The direction $\theta$ shows on the horizontal axis. The velocity $\nu$ shows on the vertical axis. The center of the bump is at $(0,0)$ shown in A1. The center of the bump shown in A2 is at $(5.68, 0.0010)$. (B, C, D) The population activity of the grid-by-velocity units. In every conjunctive grid layer, the four-dimensional population activity is sliced in the $\nu_{x}$ and $\nu_{y}$ into 49 planes, $\nu_x$ increasing from left to right and $\nu_y$ decreasing from top to bottom. In each panel, the horizontal axis is the axis $\theta_{x}$ and the vertical axis is the axis $\theta_{y}$ with the same velocity label. Three same conjunctive grid patterns are shown in B1, C1, D1 in the beginning of the experiment. With the same velocity inputs, three different conjunctive grid patterns are shown in B2, C2, D2 in the middle of the experiment. (F, G, H) The population activity of the grid units is correspondingly summed from the population activity of the conjunctive grid units along velocity axes $\nu_{x}$ and $\nu_{y}$. (E) Place firing pattern. }
\label{fig:neuralrepresentation}
\end{figure*}

Finally, we write python scripts to visualize the live state of our cognitive mapping system. In order not to trivially show the running windows of our system. The neural activity of HD-by-velocity cells, grid-by-velocity cells, grid cells and place cells can be found in Figure~\ref{fig:neuralrepresentation}. The mapping process shows all the live state of our cognitive mapping system in video S1 in Supplementary Materials, which shows the image of the scene and the local view templates, as well as the current experience map.

\section{Results of Demonstration on KITTI Datasets}

In this work, we test the Fourier hypothesis about spatial cognition and validate our entorhinal-hippocampal model based on the robotic system. Our model is demonstrated on the KITTI odometry benchmark dataset~\cite{Geiger2012CVPR}, which is recorded by a stereo camera from a car with relatively high speed in urban and highway environments. The stereo camera works at 10 Hz and records images with resolution of $1241 \times 376$ pixels. We run our cognitive mapping system on a six Intel Core i7-4930K personal computer with 64GB RAM. Video S1 in Supplementary Materials shows the mapping process of the KITTI odometry benchmark dataset sequence 00 by our implemented mapping system.

Fig.~\ref{fig:neuralrepresentation} shows two groups of activities of HD-by-velocity units (Figure~\ref{fig:neuralrepresentation}A), the grid-by-velocity units (Fig.~\ref{fig:neuralrepresentation}B, C, and D), the grid units (Fig.~\ref{fig:neuralrepresentation}F, G, and H), and the place units (Fig.~\ref{fig:neuralrepresentation}E). 
Group 1 shows activities of all units in the beginning of the experiments, when the robot is stationary. At this moment, as angular velocity is zero, the bump of HD-by-velocity units is centered in the middle of the velocity dimension shown in Fig.~\ref{fig:neuralrepresentation}A1. And since the inputs of translational velocities are also zero, three bumps of grid-by-velocity units with different frequencies are centered in the middle of velocity dimensions shown in Fig.~\ref{fig:neuralrepresentation}B1, C1, and D1. When the robot is on the original position, three grid patterns in the initial state shown in Fig.~\ref{fig:neuralrepresentation}F1, F1, and H1. Place firing patterns are generated by Bayesian integration of three grid modules with different spatial frequencies through attractor dynamics, shown in Fig.~\ref{fig:neuralrepresentation}E1.

And group 2 shows activities of all units at a randomly selected time stamp, when the robot is moving. The activity bump of HD-by-velocity is centered at (5.68, 0.0010) shown in Fig.~\ref{fig:neuralrepresentation}A2, which means the robot is heading in $325.6^\circ$ and rotating at 0.10 rad/s. The same velocity is fed into three grid-by-velocity modules, whose grid activity bumps move with the same velocity (Fig.~\ref{fig:neuralrepresentation}B2, C2, and D2). Three grid patterns with different frequencies emerge to represent the same location in the environment shown in Fig.~\ref{fig:neuralrepresentation}F2, F2, and H2. Three spatial frequencies of grid cells mold the place firing patterns through Bayesian mechanism shown in Fig.~\ref{fig:neuralrepresentation}E2.

\subsection{Place Firing Patterns Generated from Three Spatial Frequencies of Grid Modules}

In the previous subsection, the neural representation of the proposed model has been presented. Then, the place firing activity patterns formed by three spatial frequencies of grid modules would be further explained. According to Bayesian integration of two grid spatial frequencies in Eq. \ref{eq:bayesFourierIntegration}, for each grid activity bump, clustering activity bumps are first performed, and then Fourier transformations are used to estimate the center and the variance of each activity bump. As shown in the group 2 of Fig.~\ref{fig:neuralrepresentation}, considering the periodic boundary conditions, three grid activity patterns from high-frequency to low-frequency, i.e. Fig.~\ref{fig:neuralrepresentation}F2, G2, and H2, are clustered into 16, 4, 1 bumps, respectively. After estimation of Fourier transformation in Eq.~\ref{eq:bump_center_estimation_mean} and~\ref{eq:bump_center_estimation_variance}, estimation of grid activity bumps is presented in Table~\ref{table_estimation_grid_bumps}, formated by two dimensional bump centers and variances, i.e. $([\mu_x, \mu_y],[\sigma_x,\sigma_y])$. 
Every activity bump in the same grid module is regarded as the same with others, due to the periodic boundary conditions. Three homocentric activity bumps are selected from three grid modules, which are labeled in bold in Table~\ref{table_estimation_grid_bumps}, i.e. $([1.71, 6.20],[1.16,2.02])$, $([1.75, 6.17],[5.42,10.84])$,  and $([1.77, 0.34],[21.39,20.48])$. 
After Bayesian integration (Eq.~\ref{eq:place_gaussian_profile}) in Section~\ref{subsec:inputs_from_grids}, the input of place cell network (Eq.~\ref{eq:input_grid}) can be described by a two dimensional Gaussian profile. For further, through intrinsic attractor dynamics, the place firing pattern (Fig.~\ref{fig:neuralrepresentation}E2) can be given by a two dimensional Gaussian profile, namely $([1.72, 6.23],[0.92,1.57])$.

\begin{table*}[!ht]
\renewcommand{\arraystretch}{1.3}
\caption{Estimation of Grid Bumps $([\mu_x, \mu_y],[\sigma_x,\sigma_y])$}
\label{table_estimation_grid_bumps}
\centering
\begin{tabular}{c c c c c c c c c c c c c c c c c c c c}
\hline
\hline
Grid Modules & 1 & 2 & 3 & 4 & 5 & 6 & 7 & 8 \\
\hline
LayerHigh(1$\sim$8) & \shortstack{$([0.14,6.20]$, \\ $[1.16,2.02])$} & \shortstack{$\mathbf{([1.71,6.20]}$, \\ $\mathbf{[1.16,2.02])}$} & \shortstack{$([3.28,6.20]$, \\ $[1.16,2.02])$ } & \shortstack{$([4.85,6.20]$, \\ $[1.16,2.02])$}& \shortstack{$([0.92,1.48]$, \\ $[1.13,1.54])$} & \shortstack{$([2.50,1.48]$, \\ $[1.13,1.54])$} & \shortstack{$([4.07,1.48]$, \\ $[1.13,1.54])$ } & \shortstack{$([5.64,1.48]$, \\ $[1.13,1.54])$}\\
LayerHigh(9$\sim$16) & \shortstack{$([0.14,3.06]$, \\ $[1.17,2.02])$} & \shortstack{$([1.71,3.06]$, \\ $[1.16,2.02])$} & \shortstack{$([3.28,3.06]$, \\ $[1.16,2.02])$ } & \shortstack{$([4.85,3.06]$, \\ $[1.16,2.02])$}& \shortstack{$([0.93,4.62]$, \\ $[1.13,1.54])$} & \shortstack{$([2.50,4.62]$, \\ $[1.13,1.54])$} & \shortstack{$([4.07,4.62]$, \\ $[1.13,1.54])$ } & \shortstack{$([5.64,4.62]$, \\ $[1.13,1.54])$}\\
LayerMid & \shortstack{$\mathbf{([1.75,6.17]}$, \\ $\mathbf{[5.42,10.84])}$} & \shortstack{$([4.89,6.17]$, \\ $[5.42,10.84])$ } & \shortstack{$([0.18,3.03]$, \\ $[5.42,10.85])$} & \shortstack{$([3.32,3.03]$, \\ $[5.42,10.84])$}\\
LayerLow & \shortstack{$\mathbf{([1.77,0.34]}$, \\ $\mathbf{[21.39,20.48])}$}\\
\hline
\hline
\end{tabular}
\end{table*}

\subsection{Cognitive Map}


\begin{figure}[!ht]
\begin{center}
\includegraphics[width=8cm]{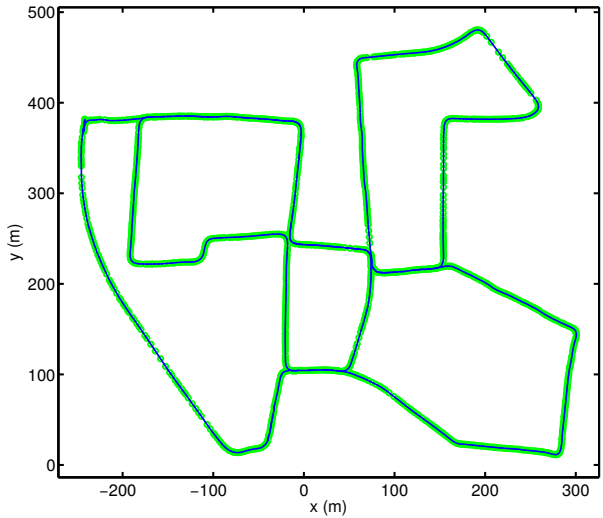} 
\end{center}
\caption{The cognitive map is a semi-metric topological map of the KITTI odometry benchmark dataset sequence 00 created by the mapping system. The topological vertices is presented by the
small green circles. The blue thin line describes links between connected vertices.}
\label{fig:kitti_cognitivemap}
\end{figure}

\begin{figure}[!ht]
\begin{center}
\includegraphics[width=8cm]{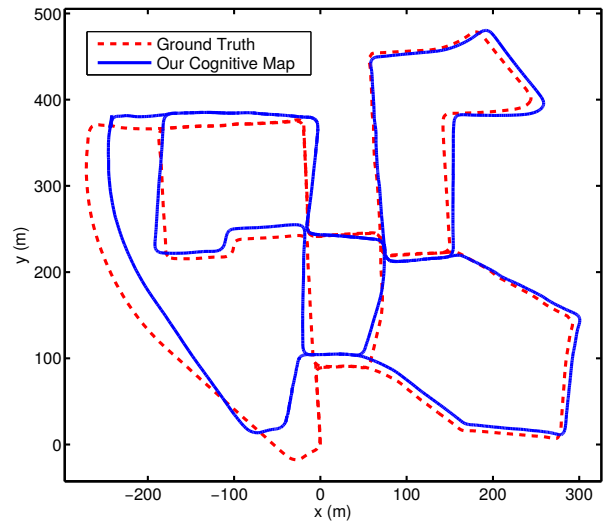} 
\end{center}
\caption{Our cognitive map (Blue) and ground truth (Red) of sequence 00 from the KITTI odometry benchmark dataset.}
\label{fig:kitti_cognitivemap_groundtruth}
\end{figure}

The cognitive map of the KITTI odometry benchmark dataset sequence 00 is generated by the implemented mapping system shown in Figure~\ref{fig:kitti_cognitivemap}. The vertices in the topological map are present by the thick green line, which represents the robot position in the explored environment. Two related vertices are connected by the link presented by fine blue line.
As the link also contains the information about the physical distance, the experience map becomes a semi-metric topological map.

The cognitive map of the KITTI odometry benchmark dataset sequence 00 is qualitatively compared with the ground truth shown in Figure~\ref{fig:kitti_cognitivemap_groundtruth}.
The cognitive map captures the overall layout of the road network, including loop closures, intersections, corners, and curves intersections, which can be clearly seen using naked eyes. On the whole, the cognitive mapping system can build the cognitive map consistent with the ground truth of the environment.

\subsection{Firing Rate Maps}

\begin{figure}[!ht]
\begin{center}
\includegraphics[width=9cm]{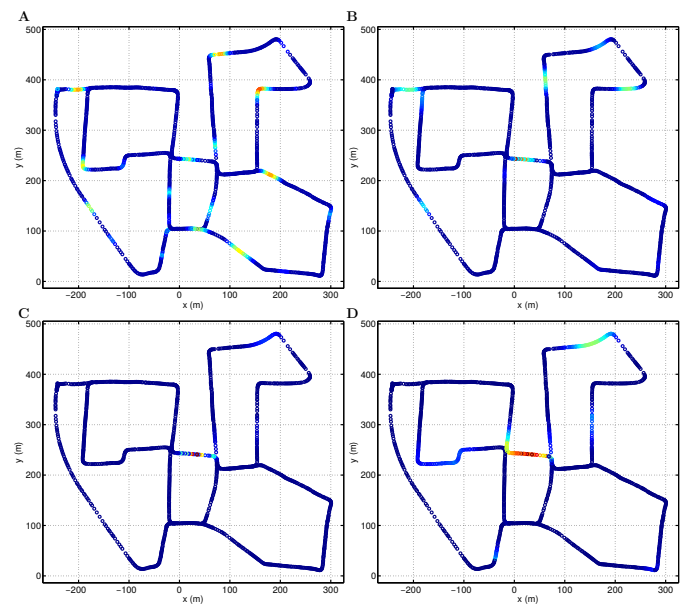} 
\end{center}
\caption{Firing rate maps of grid cells with high frequency (A), middle frequency (B), as well as low frequency (C), and place cells (D). In each panel, the firing rate is color-coded by the jet colormap from blue (zero firing rate) to red (high firing rate). }
\label{fig:kitti_firingratemaps}
\end{figure}

The place activity patterns with a single bump are generated from three grid layers with different frequencies from high to low, which encode the locations of the robot during the exploring process. Further to explore the Fourier hypothesis, firing rate maps of grid cells and place cells are shown in Fig.~\ref{fig:kitti_firingratemaps} to describe the formation of the place field. Each grid network has the same number of units as the number in the place network. The units with the same index in three grid networks and a place cell network have been selected to draw firing rate maps. Firing rate maps of grid cells shown in  Fig.~\ref{fig:kitti_firingratemaps}A, B, and C present three firing rate maps from high-frequency to low-frequency, which correspond to the firing activity grid patterns in Fig~\ref{fig:neuralrepresentation}F, G and H, respectively. For the grid units in the higher frequency, more firing fields are available for them (like Fig.~\ref{fig:kitti_firingratemaps}A and B). For grid firing field with lower frequency, the firing field is much larger than the one with higher frequency. 
The firing rate map of the place unit shown in Fig.~\ref{fig:kitti_firingratemaps}D has a similar firing field as the firing rate maps of three grid patterns, which is molded by three spatial frequencies of grid cells.

\section{Discussion}


In this work, following anatomical connectivity and neurobiological results, we further investigate the computational entorhinal-hippocampal model from grid cells to place cells. We firstly derive the model from different grid spatial frequencies to the place field to interpret the place field expansion after focal MEC inactivations~\cite{ormond_place_2015,kubie_spatial_2015} and reinterpret Fourier hypothesis by Bayesian inference, not by Fourier-like summation. Then, in order to account for the whole organization of spatial navigation and episodic memory based on the entorhinal-hippocampal model and neocortex, we propose the spatial memory indexing theory based on the hippocampal indexing theory~\cite{teyler_hippocampal_1986,teyler_hippocampal_2007}. Plus, we resort to the robot system for testing the Fourier hypothesis with Bayesian inference and the spatial memory indexing theory, and validating models of the entorhinal-hippocampal circuit. Our proposed model is implemented on a vision-only robot mapping system based on ROS, and demonstrated on the KITTI odometry benchmark dataset. Our system is capable of building a coherent semi-metric topological map in the large-scale outdoor environment (see Video S1 in Supplementary Materials). 

Recently, experiments largely, but not completely, support the Fourier hypothesis by Fourier-like summation~\cite{ormond_place_2015,kubie_spatial_2015}. As this hypothesis predicted, dorsal MEC inactivation leads to strongly broaden the spatial tuning of place cells. However, against the prediction, inactivating the ventral MEC also weakly expands the spatial firing patterns. Ormond and McNaughton suggest that the MEC inactivation leads to decrease the self-motion input to grid cells, which causes grid scale expansion and further results in place field expansion~\cite{ormond_place_2015}. Further experiments and models are required to test it.
Kubie and Fox suggest that since the discrete step size between two grid spatial frequencies is fairly large, the highest grid frequency would determine the place field size, and the lowest grid frequency has little effect on the shape of the firing field~\cite{kubie_spatial_2015}. 
To put an end to the contradiction about the prediction of Fourier hypothesis by Fourier-like summation, we propose an improved Fourier hypothesis by Bayesian inference to provide a theoretical support that place firing fields can be molded by the spatial frequencies of grid cells. The prediction of our proposed model is completely consistent with experiment results. Any parts of MEC that are inactivated would broaden the place firing fields. As the place field is largely determined by the high-frequency grid components, inactivation of dorsal   MEC causes a strong expansion of the place firing field, whereas inactivation of ventral MEC only has a weak tendency to broaden the place field. In addition, our model also gives an explanation of grid component alignment. The ventral MEC determines the rough location of the place field, namely in which high-frequency cycle, whereas the dorsal MEC determines the precise location and the shape of the place firing field. According to the above principles, even the grid patterns are not aligned very accurately, which can also be tolerant.

Based on the anatomic connections, intrinsic organization, and cognitive functions of the entorhinal-hippocampal circuit, we propose the spatial memory indexing theory upon the hippocampal indexing theory to organically incorporate the entorhinal cortex, the hippocampus, and neocortex that stores episodic memory. The essential of our idea is that when the brain experiences a new episode, the neocortex can capture information and store as the neocortical activity patterns, and the entorhinal-hippocampal cortex uses it to do path integration and then generates an index for the current episodic memory~\cite{teyler_hippocampal_2007}. Moreover, as many neurons in the primary visual cortex (V1) co-fluctuate with neurons in CA1~\cite{haggerty2015activities}, the brain experiences the familiar episode that could activate the stored neocortical activity patterns. Then the hippocampus can serve as an index to the pattern of neocortical activity, and project back to the entorhinal cortex to activate the corresponding grid firing patterns in the deep layer of MEC. Thus the stored episodic memory in the neocortex can calibrate the spatial navigation information in the entorhinal-hippocampal circuit during the exploration of the environment. 
 
Since the KITTI odometry benchmark dataset is recorded from a car in the large-scale urban environments, it is very unlikely to repeat exploration like the rat in the biological experiments, and so such hexagon firing grid patterns are not available in our firing rate maps (Fig.\ref{fig:kitti_firingratemaps}). We can also find that three grid layers with different frequencies have the same firing field, which is also found in the firing rate map of place cells. As for the firing pattern of grid cells, influenced by the accuracy of path integration by attractor network, the bump centers of three grid patterns may drift a little bit (\ref{table_estimation_grid_bumps}), namely, not always align accurately. Limited by physical memory of the computer, each grid layer only has 20-by-20 units, which leads to that the grid layer with higher frequency can not form bumps with the ideal profile (Fig.~\ref{fig:neuralrepresentation}F and G). Thus, when receiving the input from three grid frequencies (Eq.~\ref{eq:input_grid}), the place activity bumps are also slightly elliptic.

The place responses in the dentate gyrus can be generated from the layer II of MEC by self-organized plasticity~\cite{rolls2006entorhinal,si2009role}, which performs pattern separation functions. Nevertheless, to get the whole picture of episodic memory and spatial navigation, more detailed works are required to figure out relations between dentate gyrus and connected regions. 
Fourier-like summation of inputs over a range of grid spatial scales also has been proposed to explain the place responses in CA3 and CA1 regions, but recently neurobiological results can not be completely explained~\cite{solstad_grid_2006, ormond_place_2015, kubie_spatial_2015}. Based on the hypothesis of Fourier-like summation, we propose an improved Fourier hypothesis with Bayesian inference, whose prediction is consistent with the results.

Several limitations have remained in our work. First, in the entorhinal-hippocampal model, the same neural representations are used in CA3 and CA1 regions, and long-term potentiation from CA3 to CA1 is also not considered as a dynamic process. Second, to simplify our model, the dentate gyrus is not incorporated in our system. Third, the projects between CA1 and MEC, as well as CA1 and visual cortex are not represented by weights, but by corresponding relations.

For future works, we will include the cognitive function of the dentate gyrus to distinguish different scenes and break the perception of ambiguity during environment exploration. Also, we plan to build a platform, on which the robot can repeat exploration like the rat in the biological experiments, and hexagon firing grid patterns may emerge in the firing rate maps.

\section{Conclusion}

In summary, we propose an entorhinal-hippocampal model to test Fourier hypothesis and spatial memory indexing theory, and resort to the robot system to validate our proposed entorhinal-hippocampal model. The improved Fourier hypothesis in a general Bayesian mechanism predicts the place field expansion consistent with the neurobiological experimental results, and further explains the alignment of multiple grid patterns upon a range of spatial scales. This Bayesian mechanism may pertain to other neural systems as a general manner of representation, in addition to the entorhinal-hippocampal circuit. The spatial memory indexing theory is introduced from the hippocampal indexing theory to account for spatial memory and navigation during the environment exploration. Moreover, the model inspires a SLAM system able to successfully build the coherent semi-metric topological map on the benchmark dataset, and it also provides a possible opportunity to do interaction research between robotics and neuroscience.


\ifCLASSOPTIONcaptionsoff
  \newpage
\fi

\bibliographystyle{IEEEtran}
\bibliography{IEEEabrv,fourier}



\vfill


\end{document}